\documentclass[aps,prl,reprint,groupedaddress]{revtex4-1}
\usepackage{graphicx} 
\usepackage{dcolumn} 
\usepackage{bm}
\usepackage{hyperref}

\begin{document}

\title{Isospin effect in peripheral heavy-ion collisions at Fermi energies}
\author{Ya-Fei Guo$^{1,2}$, Peng-Hui Chen$^{1,2}$, Fei Niu$^{3}$, Zhao-Qing Feng$^{3}$}
\email{Corresponding author: fengzhq@scut.edu.cn}

\affiliation{$^{1}$Institute of Modern Physics, Chinese Academy of Sciences, Lanzhou 730000, People's Republic of China     \\
$^{2}$University of Chinese Academy of Sciences, Beijing 100190, People's Republic of China     \\
$^{3}$School of Physics and Optoelectronics, South China University of Technology, Guangzhou 510640, People's Republic of China  }

\date{\today}

\begin{abstract}
Within the framework of the Lanzhou quantum molecular dynamics (LQMD) transport model, the isospin effect in peripheral heavy-ion collisions has been investigated thoroughly. A coalescence approach is used for recognizing the primary fragments formed in nucleus-nucleus collisions. The secondary decay process of the fragments is described by the statistical code, GEMINI. Production mechanism and isospin effect of the projectile-like and target-like fragments are analyzed with the combined approach. It is found that the isospin migration from the high-isospin density to the low-density matter takes place in the neutron-rich nuclear reactions, i.e., $^{48}$Ca+$^{208}$Pb, $^{86}$Kr+$^{48}$Ca/$^{208}$Pb/$^{124}$Sn, $^{136}$Xe+$^{208}$Pb, $^{124}$Sn+$^{124}$Sn and $^{136}$Xe+$^{136}$Xe. A hard symmetry energy is available for creating the neutron-rich fragments, in particular in the medium-mass region. The isospin effect of the neutron to proton (n/p) ratio of the complex fragments is reduced once including the secondary decay process. However, a soft symmetry energy enhances the n/p ratio of the light particles, in particular at the kinetic energies above 15 MeV/nucleon.

\begin{description}
\item[PACS number(s)]
21.65.Ef, 24.10.Lx, 25.75.-q
\end{description}
\end{abstract}

\maketitle

\section{I. Introduction}

Over past several decades, the quasi-fission or fast fission dynamics in massive nuclear reactions attracted much attention both in experiments and in theories \cite{It15,Za07,Fe09a,Ar12}. The characteristics of the quasi-fission reactions has been extensively studied, in particular for understanding the formation mechanism of superheavy nuclei. The composite system in the quasi-fission reactions persists a long time scale of the relative energy dissipation and nucleon transfer, in which the isospin dependent nucleon-nucleon potential dominates the fragment formation. It is well known the nuclear equation of state (EOS) is a basic quantity for describing the nuclear matter with the density and temperature. The cold EOS in the isospin asymmetric nuclear matter is usually expressed with the energy per nucleon as $E(\rho, \delta)=E(\rho, \delta=0)+E_{\textrm{sym}}(\rho)\delta^{2}+ \mathcal{O}(\delta^{2})$ in terms of baryon density $\rho$ and isospin asymmetry $\delta=(\rho_{n}-\rho_{p})/(\rho_{n}+\rho_{p})$. The symmetry energy $E_{\textrm{sym}}$ at the subsaturation densities is a key ingredient in understanding the structure of weakly bound nucleus, nuclear liquid-gas phase transition, nuclear boundary of the neutron-rich nuclide etc \cite{Ch04,Co93,Po95,Ma99}. A soft symmetry energy in the domain of subsaturation densities has been extracted from the different observables in Fermi-energy (10-100 MeV/nucleon) heavy-ion collisions \cite{Ba01,Ch05,Li08,Su10,Ko11}. Different mechanisms might coexist and be correlated in heavy-ion collisions at the Fermi energies, such as projectile or target fragmentation, neck dynamics and fragment formation, preequilibrium emission of light clusters (complex particles), fission of heavy fragments, multifragmentation etc, in which the isospin dependent nucleon-nucleon potential dominates the dynamical processes. Recently, the isospin effects of the intermediate mass fragments (IMFs) from the neck fragmentation were investigated thoroughly \cite{Fi12,Fi14,Fe16}.

The observables in the quasi-fission or fast fission reactions are related to the isospin dependent mean-field potential and have the long isospin relaxation time. As a typical quantity, the neutron-to-proton ratios ($N/Z$) of the fragments reflects the isospin diffusion in nucleus-nucleus collisions. In this work, we investigate the isospin dynamics of the fragment formation and decay modes within the Lanzhou quantum molecular dynamics (LQMD) transport model. In Sec. II we give a brief description of the LQMD model. The isospin dissipation dynamics and the symmetry energy effect are discussed in Sec. III. Summary and perspective on the possible measurements are presented in Sec. IV.

\section{II. Brief description of the LQMD model}

To understand the isospin dynamics in the medium and high energy heavy-ion collisions, the LQMD transport model is developed on the issues of symmetry energy at subnormal and suprasaturation densities, in-medium nucleon-nucleon (NN) cross sections, isospin and momentum dependent NN potential, isospin effects on particle production and transportation etc \cite{Fe11,Fe14,Fe15,Fe18}. All possible reaction channels of the charge-exchange, elastic and inelastic scatterings by distinguishing isospin effects in hadron-hadron collisions were implemented in the model. The temporal evolutions of the baryons (nucleons and resonances) and mesons in the colliding system under the self-consistently generated mean-field are described by Hamilton's equations of motion. The Hamiltonian is constructed with the Skyrme-type interaction, which consists of the relativistic energy and the effective interaction potential. We do not take into account the momentum-dependent potential in the Fermi-energy and near barrier nuclear collisions for stabilizing the initial nucleus. However, the momentum-dependent interaction has to be included in the high-energy heavy-ion collisions, which reduces the effective nucleon mass in nuclear medium.

The local potential can be written with the energy-density functional as
$U_{loc}=\int V_{loc}(\rho(\mathbf{r}))d\mathbf{r}$. The energy-density functional is expressed by
\begin{eqnarray}
V_{loc}(\rho)=&& \frac{\alpha}{2}\frac{\rho^{2}}{\rho_{0}} +
\frac{\beta}{1+\gamma}\frac{\rho^{1+\gamma}}{\rho_{0}^{\gamma}} + E_{sym}^{loc}(\rho)\rho\delta^{2}
\nonumber \\
&& + \frac{g_{sur}}{2\rho_{0}}(\nabla\rho)^{2} + \frac{g_{sur}^{iso}}{2\rho_{0}}[\nabla(\rho_{n}-\rho_{p})]^{2},
\end{eqnarray}
where $\rho_{n}$, $\rho_{p}$ and $\rho=\rho_{n}+\rho_{p}$ are the neutron, proton and total densities, respectively, and  $\delta=(\rho_{n}-\rho_{p})/(\rho_{n}+\rho_{p})$ being the isospin asymmetry. The parameters $\alpha$, $\beta$, $\gamma$, $g_{sur}$ $g_{sur}^{iso}$ and $\rho_{0}$ are taken to be -226.5 MeV, 173.7 MeV, 1.309, 23 MeV fm$^{2}$, -2.7 MeV fm$^{2}$ and 0.16 fm$^{-3}$, respectively. These parameters lead to the compression modulus of K=230 MeV for isospin symmetric nuclear matter at saturation density. The $E_{sym}^{loc}$ is the local potential of the symmetry energy given by
\begin{eqnarray}
E_{sym}^{loc}(\rho)=\frac{1}{2}C_{sym}(\rho/\rho_{0})^{\gamma_{s}}
\end{eqnarray}
where the coefficients $\gamma_{s}$=0.5, 1 and 2 correspond to the soft, linear and hard symmetry energies, respectively. The value of $C_{sym}$=38 MeV results in the symmetry energy of 31.5 MeV at the normal nuclear density.

In the low-energy heavy-ion collisions, the fermionic nature of nucleons and NN elastic scattering dominate the nuclear dynamics and the fragment formation. The treatment of the Pauli blocking in NN collisions is different in various transport models \cite{Zh18}. The purpose is to reduce the collision probability for the quantal many-body system. In the LQMD model, the Pauli blocking in NN collisions is embodied by the blocking probability $b_{ij}=1-(1-f_{i})(1-f_{j})$ in two colliding nucleons $i$ and $j$ compared with a random number. The isospin effect is included in evaluation of the occupation probability by
\begin{eqnarray}
f_{i}=  && \frac{32\pi^{2}}{9h^{3}}\sum_{i\neq k, \tau_{i}=\tau_{k}} (\triangle r_{ik})^{2}(3R_{0}-\triangle r_{ik})  \nonumber \\
&&  (\triangle p_{ik})^{2}(3P_{0}-\triangle p_{ik}).
\end{eqnarray}
The $r_{ik}=|\mathbf{r}_{i}-\mathbf{r}_{k}|$ and $p_{ik}=|\mathbf{p}_{i}-\mathbf{p}_{k}|$ are the relative distance of two nucleons in coordinate and momentum spaces, respectively. The sum is satisfied to the conditions of $r_{ik}<2R_{0}$ and $p_{ik}<2P_{0}$ with $R_{0}=3.367$fm and $P_{0}=112.5$ MeV/c. It is noticed that the overlap of hard sphere in phase space is assumed by $16\pi^{2}R_{0}^{3}P_{0}^{3}/9=h^{3}/2$. The cross section of the NN elastic scattering is obtained by fitting the available experimental data in the wide energy region \cite{Fe09}.

The nuclear dynamics in the Fermi-energy heavy-ion collisions is described by the LQMD model. The primary fragments are constructed in phase space with a coalescence model, in which nucleons at freeze-out are considered to belong to one cluster with the relative momentum smaller than $P_{0}$ and with the relative distance smaller than $R_{0}$ (here $P_{0}$ = 200 MeV/c and $R_{0}$ = 3 fm) \cite{Fe16}. The primary fragments are highly excited. The de-excitation of the fragments is described within the GEMINI code \cite{Ch88}.

\section{III. Results and discussion}

\begin{figure*}
\includegraphics[width=15 cm]{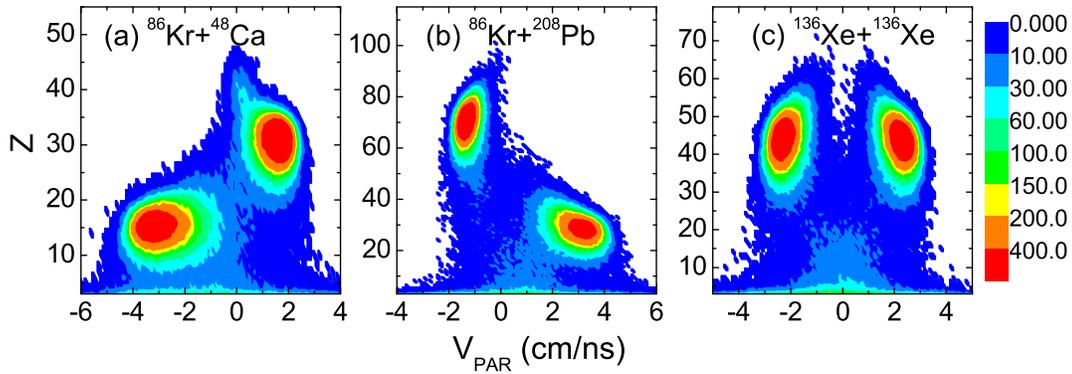}
\caption{\label{fig:wide} Fragment yields produced in collisions of $^{86}$Kr$+^{48}$Ca, $^{86}$Kr$+^{208}$Pb and $^{136}$Xe$+^{136}$Xe at the incident energy of 20 MeV/nucleon as functions of atomic number $Z$ and parallel velocity $V_{PAR}$.}
\end{figure*}

The phase-space structure of fragments produced in heavy-ion collisions provides the correlation and fluctuation of nuclear dynamics in the dissipation of relative motion energy, nucleon transfer and isospin density. The yields of the primary fragments in the peripheral HICs almost exhibit a symmetric distribution in the projectile-like fragments (PLFs) or target-like fragments (TLFs) owing to the same stripping and pick-up reaction rates. The fragment spectra are influenced by both the dynamics mechanism and structure quantities. The proton and neutron drifts in the different density region are driven by the mean-field potential in nuclear medium. Shown in Fig. 1 is the fragment distributions in the reactions of $^{86}$Kr+$^{48}$Ca/$^{208}$Pb and $^{136}$Xe$+^{136}$Xe at the incident energy of 20 MeV/nucleon in the peripheral collisions. The parallel velocity $V_{PAR}$ is evaluated along the beam direction in the center of mass frame and the positive (negative) values corresponds to the PLFs (TLFs). It is obvious that a number of fragments are created in the vicinity of projectile or target velocity. The intermediate mass fragments (IMFs) are mainly produced in the fragmentation reactions. The N/Z ratio of IMFs from the neck fragmentation in Fermi-energy heavy-ion collisions was investigated for extracting the density dependent symmetry energy \cite{Fi12,Fi14,Fe16}. Shown in Fig. 2 is the total kinetic energy (TKE) distribution of the primary fragments in the reaction of $^{48}$Ca+$^{208}$Pb within the collision centrality of 8-10 fm. The primary fragments are constructed in phase space with a coalescence model, in which nucleons at freeze-out are considered to belong to one cluster with the relative momentum smaller than $P_{0}$ and with the relative distance smaller than $R_{0}$ (here $P_{0}$ = 200 MeV/c and $R_{0}$ = 3 fm) \cite{Fe16}. It is obvious that the TKE spectra exhibit a symmetric structure. The multinucleon transfer mechanism is available for creating the medium-mass fragments with the charged numbers of Z=30-60. The dissipation of relative motion energy and nucleon transfer is more pronounced with decreasing the incident energy.

\begin{figure*}
\includegraphics[width=15 cm]{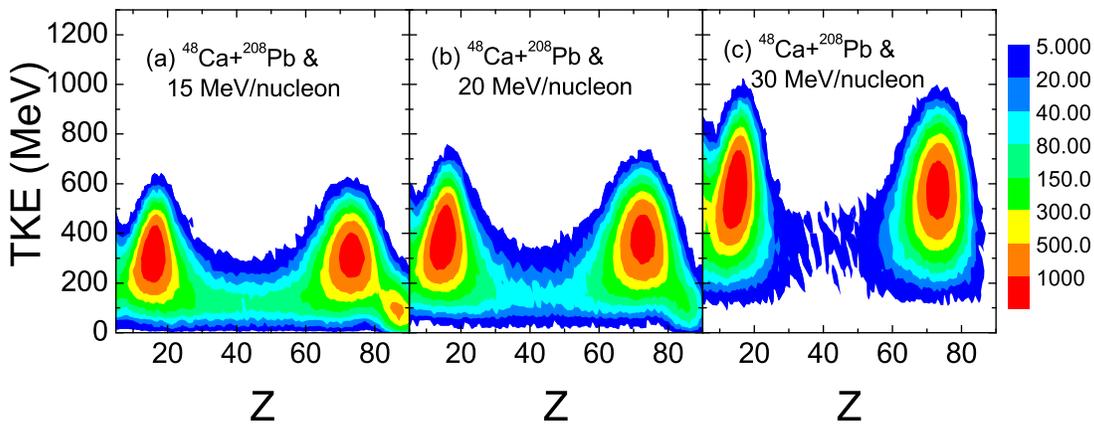}
\caption{\label{fig:wide} The total kinetic energy spectra in the $^{48}$Ca$+^{208}$Pb reaction at the energies of 15, 20 and 30 MeV/nucleon, respectively.}
\end{figure*}

\begin{figure*}
\includegraphics[width=15 cm]{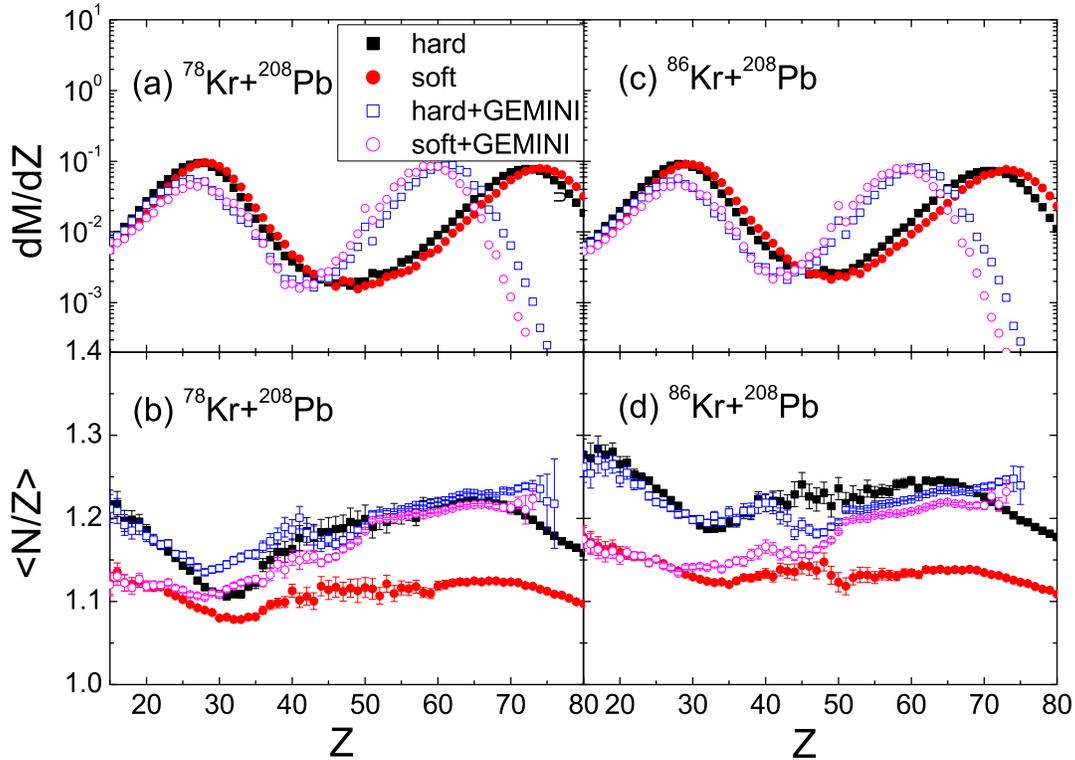}
\caption{\label{fig:wide} The total multiplicities and neutron to proton ratios as a function of the charged numbers of fragments produced in the reactions of $^{78,86}$Kr$+^{208}$Pb at 20 MeV/nucleon.}
\end{figure*}

\begin{figure}
\includegraphics[width=15 cm]{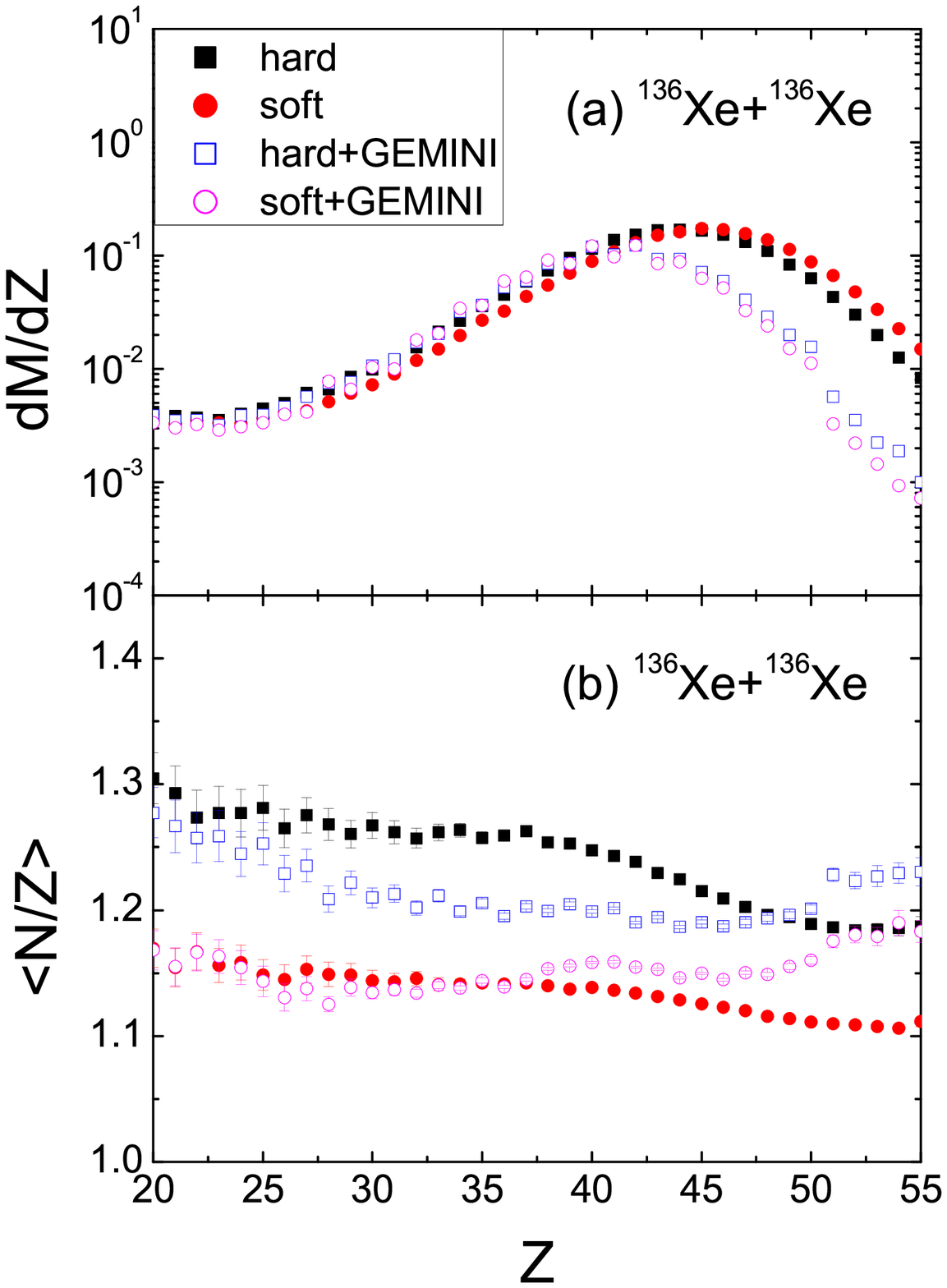}
\caption{\label{fig:epsart} The same as in Fig. 3, but for the symmetric reaction $^{136}$Xe$+^{136}$Xe at 20 MeV/nucleon with the hard and soft symmetry energies, respectively.}
\end{figure}

The isospin dynamics of fast nucleons and light complex particles produced in Fermi-energy heavy-ion collisions has been investigated for extracting the symmetry energy at subsaturation densities and the isospin splitting of nucleon effective mass in nuclear matter \cite{Gu17}. The pre-equilibrium nucleon emission in isotopic reaction systems was measured at the National Superconducting Cyclotron Laboratory (Michigan State University, East Lansing, MI, USA) \cite{Co16}. The production of fragments is calculated within the combined approach of the LQMD transport model for the primary fragments and the GEMINI statistical decay code for the de-excitation process \cite{Ch88}. The de-excitation of fragments with the charge number Z$>$6 is considered. The light fragments are only counted from the preequilibrium emission in nuclear collisions. The excitation energy with Z$>$6 is evaluated from the binding energy difference of the primary fragment and the ground state evaluated from the liquid drop model. Shown in Fig. 3 is the total multiplicities and neutron/proton ratios of heavy fragments produced in the peripheral collisions of $^{78,86}$Kr + $^{208}$Pb at the beam energy of 20 MeV/nucleon. The isospin effect is weakened via the decay process, in particular in the heavy mass domain owing to larger excitation energy. It is pronounced a hard symmetry energy is favorable for the formation of neutron-rich isotopes. However, the yields are weakly influenced the stiffness of symmetry energy. The de-excitation of primary fragments moves the spectra to the light mass region. It has been investigated that the peripheral nucleus-nucleus collisions have the advantage for creating the neutron-rich isotopes \cite{Mi11}. Shown in Fig. 4 is a comparison of different symmetry energy on the fragment distribution in the symmetric reaction $^{136}$Xe+$^{136}$Xe at the incident energy of 20 MeV/nucleon. The 'inverse quasi-fission process' leads to the one-bump structure. Moreover, the fragment yields are almost independent the stiffness of symmetry energy. However, the average neutron/proton ratios of isotopic fragments exhibit an obvious isospin effect in the wide mass region. The conclusions are helpful for extracting the symmetry energy from the quasi-fission fragments at INFN-LNS Superconducting Cyclotron of Catania (Italy) \cite{Gn17}.

\begin{figure*}
\includegraphics[width=15 cm]{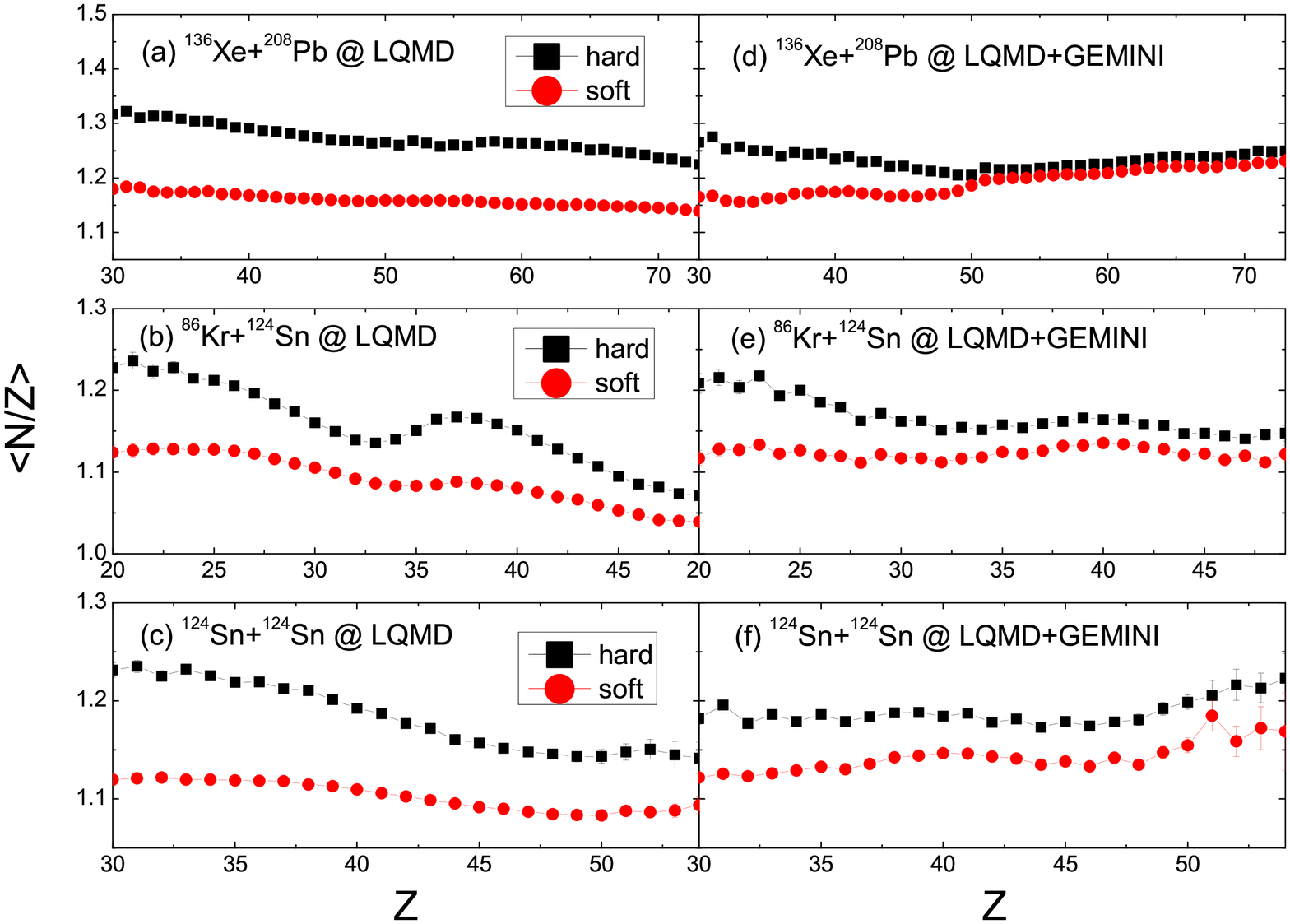}
\caption{\label{fig:wide} Comparison of the n/p ratios of the primary (left panels) and secondary (right panels) fragments with the hard and soft symmetry energies in the reactions of $^{136}$Xe$+^{208}$Pb, $^{86}$Kr$+^{124}$Sn and $^{124}$Sn$+^{124}$Sn at the energy of 20 MeV/nucleon, respectively.}
\end{figure*}

\begin{figure*}
\includegraphics[width=14 cm]{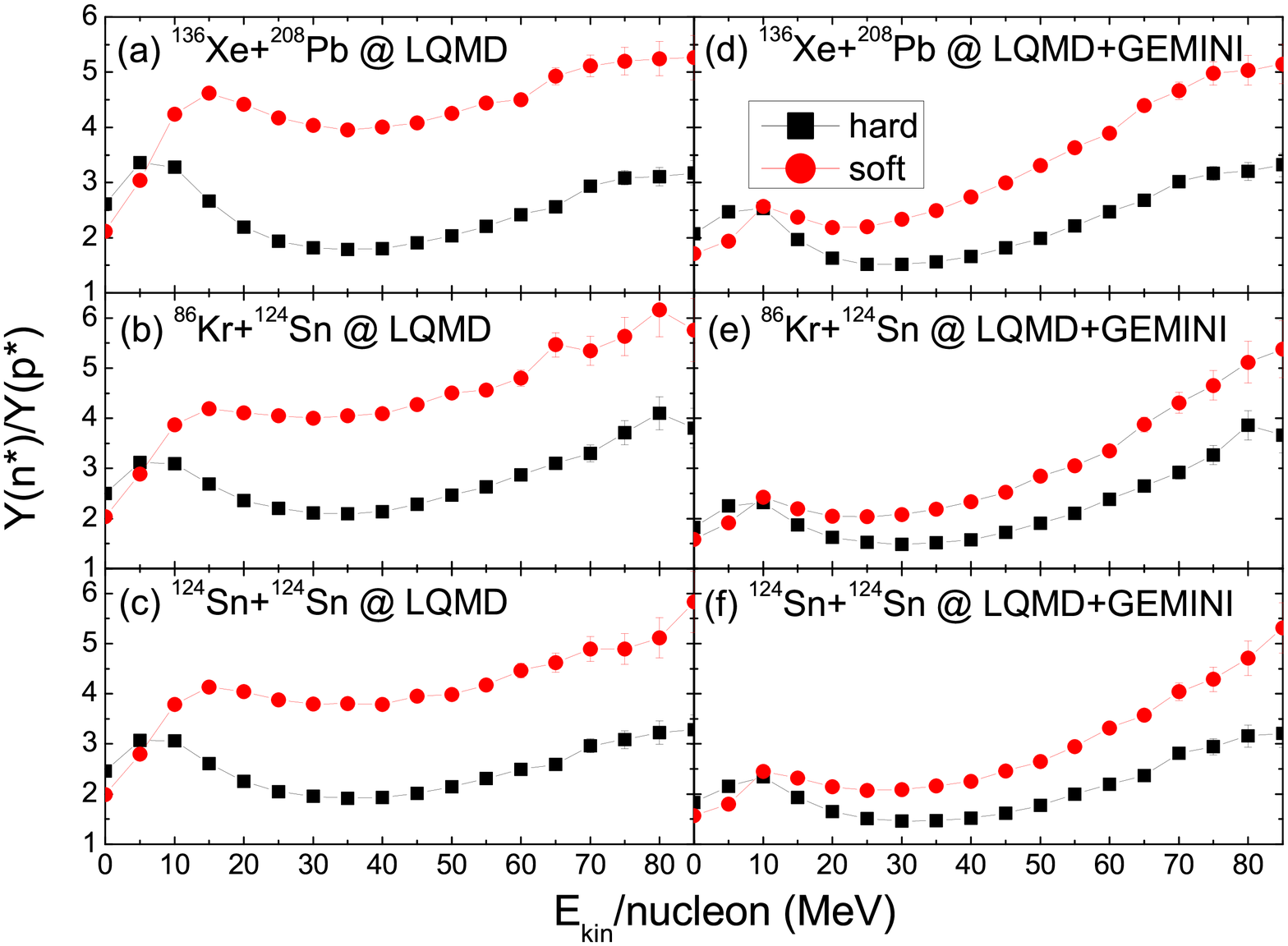}
\caption{\label{fig:wide} The same as in Fig. 5, but for the kinetic energy spectra from the yields of gas-phase particles (nucleons, hydrogen and helium isotopes).}
\end{figure*}

Systematic comparison of the n/p ratios of the primary and secondary fragments from different reaction systems is shown in Fig. 5. The isospin effect is obvious in the spectra of primary fragments, which directly brings the mean-field information. The hard symmetry energy results in the larger n/p ratio in the fragment formation because of the weaker repulsive force for neutrons in the neutron-rich matter, which is available for the neutron-rich isotope formation in peripheral heavy-ion collisions. To realize the cold fragments observed in experiments, we implemented the GEMINI code into the primary fragments. The isospin effect is weakened via the de-excitation process, in particular in the heavy-mass domain. The neutron-rich isotopes have smaller neutron separation energies. The statistical decay reduces the $\langle N/Z\rangle$ with the hard symmetry energy. However, the primary proton-rich fragments formed with the soft symmetry energy might be decayed via the proton or neutron emission. Shown in Fig. 6 is the kinetic energy spectra of the neutron/proton ratios from the 'gas-phase' fragments (nucleons, hydrogen and helium isotopes) produced in collisions of $^{136}$Xe+$^{208}$Pb, $^{86}$Kr+$^{124}$Sn and $^{124}$Sn$+^{124}$Sn at the energy of 20 MeV/nucleon, respectively. The symmetry energy effect is more pronounced for the neutron-rich system. The bump structure of the primary fragments around the energy of 15 MeV/nucleon is caused from the competition of free nucleons and light fragments to the n/p ratios. The light fragments with Z$\leq$2 are mainly produced at the low kinetic energy and have the smaller n/p ratios in comparison to the free nucleons. At the kinetic energies above 30 MeV/nucleon, the n/p ratios are mainly contributed from the free nucleons. The difference of the hard ($\gamma_{s}$=2) and soft ($\gamma_{s}$=0.5) symmetry energies is obvious in the whole energy range. Opposite trend to the $\langle N/Z\rangle$ ratio of IMFs and heavy fragments, the soft symmetry energy enhances the yield ratios because of more repulsive force for neutrons in dynamical evolution.

\begin{figure*}
\includegraphics[width=14 cm]{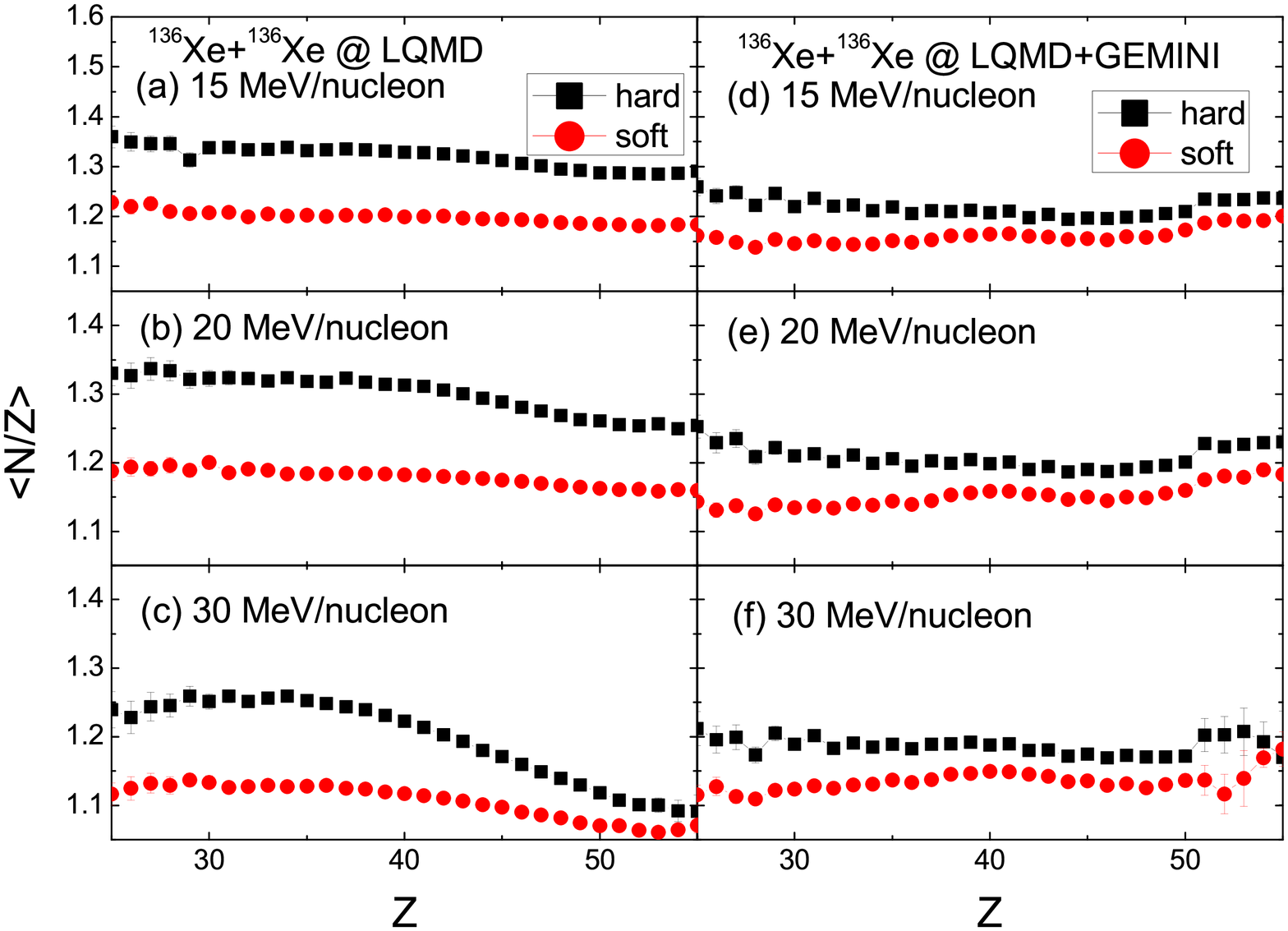}
\caption{\label{fig:wide}(Color online) Incident energy dependence of the n/p ratio in the reaction of $^{136}$Xe + $^{136}$Xe.}
\end{figure*}

\begin{figure*}
\includegraphics[width=14 cm]{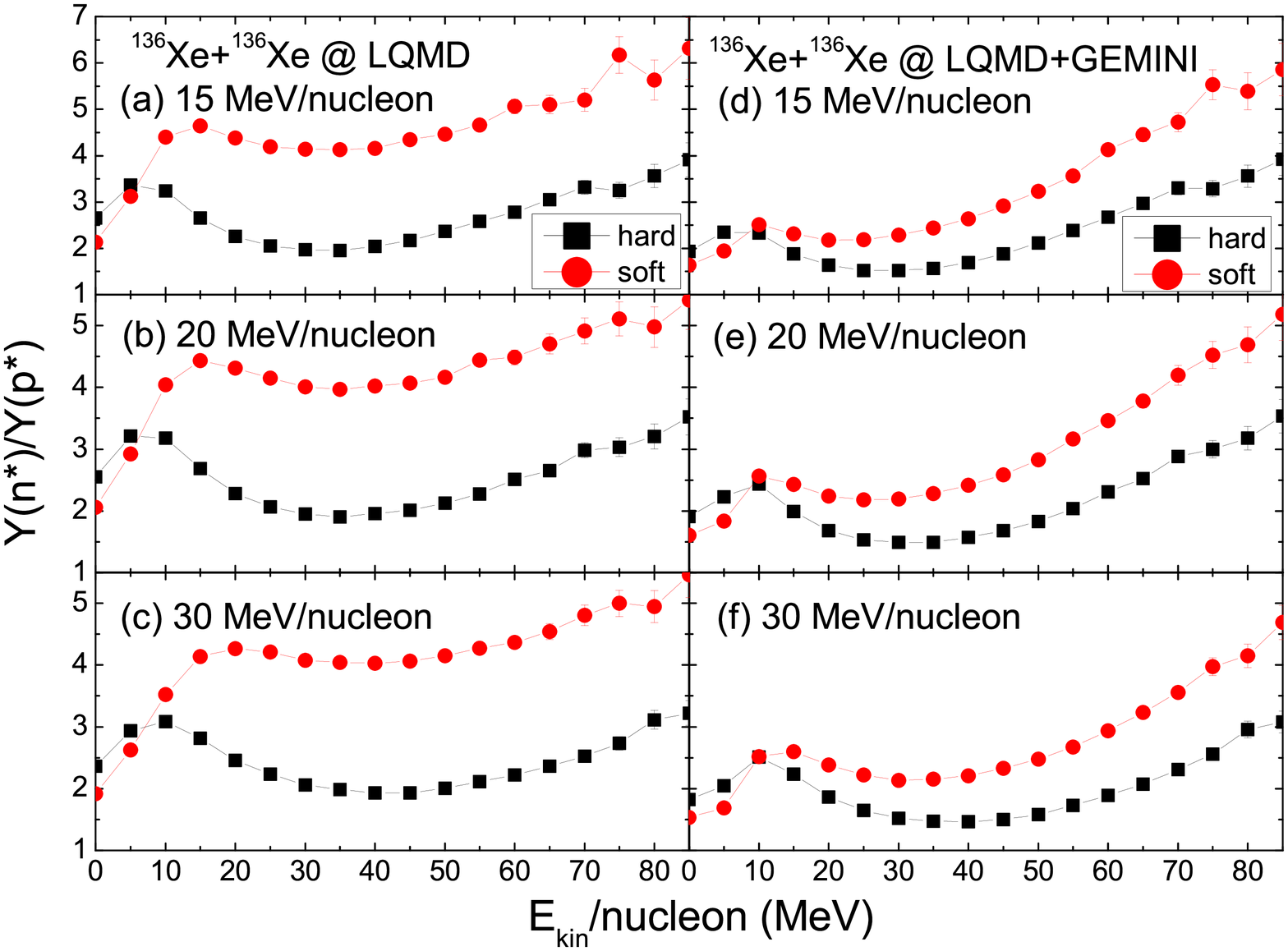}
\caption{\label{fig:wide}(Color online) The 'gas-phase' n/p ratio as a function of kinetic energy in collisions of $^{136}$Xe + $^{136}$Xe at the beam energies of 15, 20 and 30 MeV/nucleon, respectively.}
\end{figure*}

The isospin dissipation in peripheral heavy-ion collisions is related to the incident energy. The composite system maintains more temporal relaxation with decreasing the energy, in which the isospin-dependent single-particle potential impacts the nucleon dynamics and the fragment formation. Shown in Fig. 7 is a comparison of different energy on the $\langle N/Z\rangle$ ratios of fragments produced in the $^{136}$Xe + $^{136}$Xe reaction. A large $\langle N/Z\rangle$ value is obtained with the hard symmetry energy at 15 MeV/nucleon. The isospin effect is still pronounced after including the statistical decay. The dependence of the kinetic energy spectra of 'gas-phase' n/p ratio on the incident energy is small as shown in Fig. 8. The correlation measurement of the $\langle N/Z\rangle$ ratios from the heavy fragments and the n/p yield ratios from the 'gas-phase' particles is helpful in constraining the symmetry energy at subsaturation densities and the synthesis of new neutron-rich isotopes. The multinucleon transfer reaction in Fermi-energy heavy-ion collisions has been investigated within the improved quantum molecular dynamics (ImQMD) model. A narrower angular distribution of PLFs and TLFs was obtained in comparison with the low-energy nuclear collisions \cite{Ya17}.

\section{IV. Conclusions}

Within the LQMD transport model, we have investigated the isospin dynamics in the peripheral heavy-ion collisions at Fermi energies. The n/p spectra from the gas-phase particles and from the PLFs and TLFs are analyzed thoroughly. The average n/p ratios of the IMFs and heavy fragments are sensitive to the stiffness of symmetry energy because of the existence of isospin density gradient in dissipative collisions. The hard symmetry energy is available for producing the neutron-rich fragments. However, the isospin effect is reduced after implementing the statistical decay owing to the binary fission and the light particle evaporation. The larger n/p yield ratios of the 'gas phase' particles are obtained with the soft symmetry energy because of the repulsive interaction enforced on neutrons in the region of sub-saturation densities.
Further experiments are expected for extracting the subnormal symmetry energy from the medium fragments.

This work was supported by the National Natural Science Foundation of China under Grant (Projects No. 11722546 and No. 11675226), and the Major State Basic Research Development Program in China (Projects No. 2014CB845405 and No. 2015CB856903).

\end{document}